# Anisotropies of the g-factor tensor and diamagnetic coefficient in crystal-phase quantum dots in InP nanowires

Shiyao Wu[1,2], Kai Peng[1,2], Sergio Battiato[3], Valentina Zannier[3], Andrea Bertoni[4], Guido Goldoni[4,5], Xin Xie[1,2], Jingnan Yang[1,2], Shan Xiao[1,2], Chenjiang Qian[1,2], Feilong Song[1,2], Sibai Sun[1,2], Jianchen Dang[1,2], Yang Yu[1,2], Fabio Beltram[3], Lucia Sorba[3], Ang Li[6], Bei-bei Li[1], Francesco Rossella[3], and Xiulai Xu[1,2,7*]

[1] *Beijing National Laboratory for Condensed Matter Physics, Institute of Physics, Chinese Academy of Sciences, Beijing 100190, China*
[2] *CAS Center for Excellence in Topological Quantum Computation and School of Physical Sciences, University of Chinese Academy of Sciences, Beijing 100049, China*
[3] *Laboratorio NEST, Scuola Normale Superiore and Istituto Nanoscienze-CNR, Piazza S. Silvestro 12, I-56127 Pisa, Italy*
[4] *S3, Istituto Nanoscienze-CNR, Via Campi 213/a, 41125 Modena, Italy*
[5] *Dipartimento di Scienze Fisiche, Informatiche e Matematiche, Universit`a degli Studi di Modena e Reggio Emilia, Via Campi 213/a, 41125 Modena, Italy*
[6] *Beijing Key Lab of Microstructure and Property of Advanced Materials, Beijing University of Technology, Pingleyuan No.100, 100024, Beijing, P. R. China*
[7] *Songshan Lake Materials Laboratory, Dongguan, Guangdong 523808, China*
*email: xlxu@iphy.ac.cn

**Abstract**

Crystal-phase low-dimensional structures offer great potential for the implementation of photonic devices of interest for quantum information processing. In this context, unveiling the fundamental parameters of the crystal phase structure is of much relevance for several applications. Here, we report on the anisotropy of the g-factor tensor and diamagnetic coefficient in wurtzite/zincblende (WZ/ZB) crystal-phase quantum dots (QDs) realized in single InP nanowires. The WZ and ZB alternating axial sections in the NWs are identified by high-angle annular dark-field scanning transmission electron microscopy. The electron (hole) g-factor tensor and the exciton diamagnetic coefficients in WZ/ZB crystal-phase QDs are determined through micro-photoluminescence measurements at low temperature (4.2 K) with different magnetic field configurations, and rationalized by invoking the spin-correlated orbital current model. Our work provides key parameters for band gap engineering and spin states control in crystal-phase low-dimensional structures in nanowires.

## 1  Instruction

Quantum wells and quantum dots (QDs) realized in semiconductor nanowires

(NWs) represent a promising platform to implement quantum computation and quantum information processing[1,2] and this specific implementation, NWs, provides a versatile material system to demonstrate innovative nanoscale devices, such as lasers, photo-sensors and solar cells [3-9]. In this frame, compared to other semiconductor NW systems, InP NWs display a lower susceptibility to non-radiative surface states and a stronger photoluminescence (PL) emission: this material system is therefore of particular interest for quantum-photonics application [10-17]. InP NWs embedding a mixture of both WZ and ZB structures, often referred to as crystal-phase QDs, can be exploited for applications in electronics and quantum photonics, such as the generation of single photons and cascade-photon pairs [18, 19]. Recent developments of nano-fabrication technology yielded high controllability of crystal-phase NWs [20, 21] and enabled the control of quantum-state coupling and entanglement that is crucial for future applications in optics- and photonics-based quantum information processing [22].

In order to control spin properties and drive spin resonance in WZ/ZB crystal-phase QD systems, the g-factor tensor are highly desirable [23-25]. Recently, spin manipulation based on the modification of the g-factor tensor by material composition and electric field were achieved [26-28], thus showing the feasibility of spin-based quantum-information architectures based on the anisotropy of the g-factor tensor. However, to the best of our knowledge, the g-factor tensor of crystal-phase QDs implemented in InP NWs was never been reported: this lack of information poses severe limitations to the exploitation of the spin tunability and to the successful modeling and design of these low-dimensional structures.

In this work, we investigate the g-factor tensor of WZ/ZB crystal-phase QDs formed in InP NWs by carrying out angle-dependent magneto-photoluminescence (PL) experiments with a vector magnetic field system at 4.2 K. The WZ and ZB crystal structures in InP NWs are characterized by the high-angle annular dark-field (HAADF) imaging from the scanning transmission electron microscopy (STEM). The excitonic emission of the WZ/ZB crystal-phase QDs in single InP NWs are investigated in the presence of an applied magnetic field in-plane and out-of-plane with respect to the surface on which the NWs are deposited. The optically-allowed states (bright states) and -forbidden states (dark states) of the negatively-charged-exciton in crystal-phase InP QDs display Zeeman splitting and diamagnetic shifts. The anisotropies of the electron and hole g-factor tensor and the exciton diamagnetic coefficients resulting from the structural anisotropy are quantitatively addressed. The experimental results are rationalized in the frame of the spin-correlated orbital current model. Our findings provide crucial parameters for the exploitation of crystal-phase QDs in quantum devices, and pave the way to a novel approach for controlling the spin properties, such as controlling the spin precession or driving spin resonance by modifying the g-factor tensor.

## 2 Results and discussion
### 2.1 WZ/ZB crystal-phase structure in InP NWs

Figure 1 (a) shows the dark-field transmission electron microscopy (TEM) image of a single InP NW. Typically, NWs display a length of 700-800 nm with a diameter varying from 30 to 80 nm from the tip to the base of the NW. The tapered morphology is usually observed in InP NW grown by CBE and MOCVD at low growth temperature and low V/III ratio [29, 30]. Indeed, under these growth conditions adatom diffusion length is short and there is a non-negligible radial growth at the NW base simultaneously occurring with the NW axial growth. The resulting tapering is not detrimental for the NW optical properties [19, 30-33]. The HAADF measurements of different sections of the InP NW are shown in Fig. 1 (b). The WZ and the ZB crystal segments are marked with orange and green lines. In our InP NWs, the ZB segments are predominant with a few WZ segments ranging from 2 to more than 10 monolayers, forming three types of InP crystal-phase structure configurations. One first configuration comprises a long ZB segment with a rotationally twinned (RT) structure that usually introduces one unit of the WZ crystal structure [34]. This configuration occurs in the area marked with the blue square in Fig. 1 (b). The atomic STEM images with ball and stick model of this RT structure are shown in Fig. 1 (c). The other two configurations labelled by orange triangles (green dots) in Fig. 1 (b) are a WZ (ZB) segment surrounded by two ZB (WZ) segments, with the corresponding ball and stick models and atomic STEM images shown in Fig. 1 (d) and Fig. 1 (e), respectively.

Compared with WZ, the ZB structure has a smaller band gap (Fig. 1 (f)) [35, 36]. Its conduction and valence band extrema are about 129 meV and 45 meV lower than WZ structures, respectively, leading to a type-II band alignment of adjacent WZ and ZB segments and the formation of WZ/ZB crystal-phase QDs [8, 17, 19, 20]. Figure 1 (f) shows a schematic diagram of the possible exciton radiative-recombination pathways in WZ/ZB crystal-phase QDs with electrons and holes confined in the ZB and WZ segments, respectively. Here the blue and green colored areas schematically shown in Fig. 1 (f) represent the different WZ/ZB crystal-phase QDs with different size and consequently different quantization energies due to the length variation of the WZ and ZB segments (see Fig. 1 (b)) [37], resulting in different recombination spectra, as observed through micro-PL measurements.

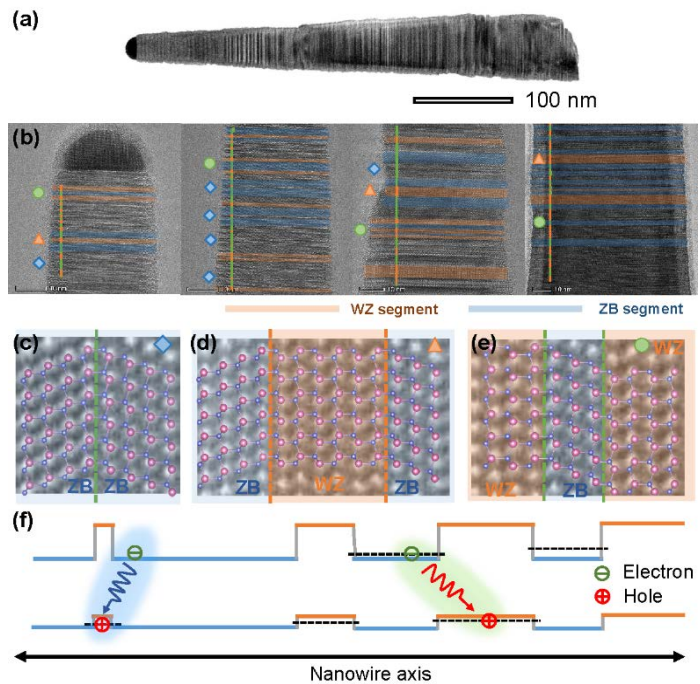

**Figure 1** Structure of InP NWs: (a) Dark field STEM image of one single representative InP NW. (b) STEM-HAADF images of different InP NW segments. The WZ and ZB segments are identified with orange and green lines, respectively. Here, the different symbols correspond to the three different structures depicted in Fig. (c)-(e), where the ball-and-stick models of the RT ZB segment, ZB/WZ/ZB and WZ/ZB/WZ structures and atomic resolution STEM images are reported. (f) The schematic energy level of the WZ/ZB sequence in the InP NW. The blue and green colored areas represent the possible exciton radiative-recombination in crystal-phase QDs with different lengths of the WZ and ZB segments.

## 2.2 Micro-PL spectroscopy

Figure 2 (a) shows the schematic diagram of the setup for μ-PL and magneto-PL measurements. Non-resonant excitation was achieved by using a laser at 650 nm. Emission from the NWs was detected by a spectrometer and a charge coupled device camera. Figure 2 (b) shows the PL spectra of five individual InP NWs. In each one, emission from InP NW as well as from the embedded crystal-phase QDs can be identified. The broad emission peaks at the high-energy side of the spectra are ascribed to the e-h recombination within InP NW. The multi-peaks with narrow linewidth at lower energy correspond to the emissions from the crystal-phase QDs formed in the NW. The PL peak energies of the crystal-phase QDs ranging from 1.27 to 1.35 eV depend on the electron (hole) quantization energy levels in crystal-phase QDs that vary owing to the variation of the ZB (WZ) segment lengths [37], where the length of ZB segments are ranging from 2 to more than 10 monolayers given by HAADF measurements. Although, we cannot identify the exact size of the crystal phase quantum dot in single InP NW for each peak as our laser spot size is much larger than the segment length. Nevertheless, the isolated single peaks from single crystal phase QDs are clear, which does not affect the g factor and diamagnetic effect analysis in this work. Figure

2 (c) shows the power-dependent PL spectra of one individual InP NW. At low excitation powers, emission from the QDs stems from the occupation of the low energy levels. As the excitation power increases, the occupation of levels of increasing energy promotes the emission from InP NW, at the high energy side. Consistently, the emission from InP NW displays a blue shift with increasing excitation power. The weakness or missing of the emission peaks with power increasing is due to the competition between QDs in NW or PL quenching for single dot by high excitation power, which is normal for single quantum dots at high excitation power.

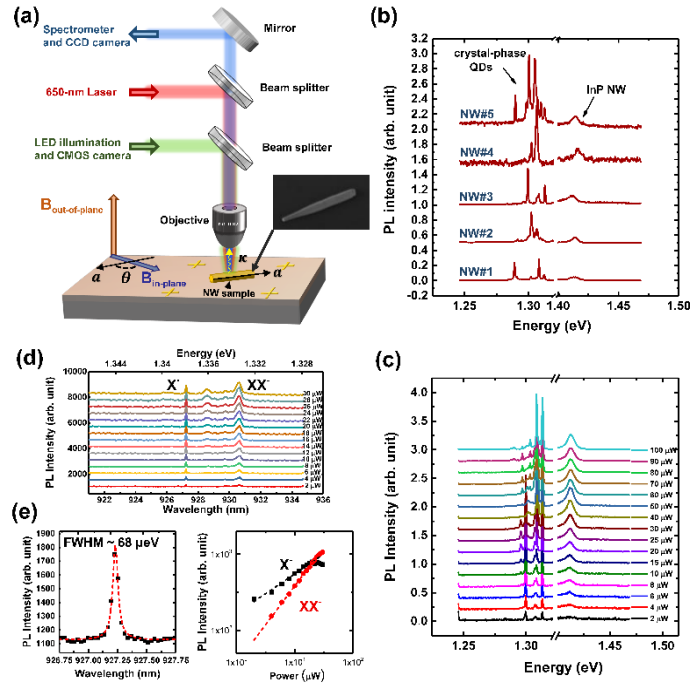

**Figure 2** (a) The schematic diagram of the setup for the μ-PL and magneto-PL measurements. Inset: SEM micrograph of one representative individual InP NW on Si/SiO$_2$ substrate. (b) Normalized PL spectra of five single InP NWs. (c) The PL spectra of one single InP NW under different excitation powers. (d) The power-dependent PL spectra of crystal-phase QDs in single NW. The two emission peaks are identified as negatively-charged-exciton (X$^-$) and negatively-charged-biexciton (XX$^-$). The spectra are shifted for clarity. (e) Left panel: PL spectrum of X$^-$ with line width of 63 μeV. Right panel: The PL intensity of X$^-$ and XX$^-$ as a function of excitation power, the circles (squares) are experimental data and the dashed black (red) line are the guide for nearly linear (quadratic) dependence.

InP NWs grown by chemical beam or molecular beam epitaxy are usually n-type with doping density in the range of $10^{17}$ and $10^{15}$ cm$^{-3}$ [38]. This provides excess electrons in InP NWs leading to the dominance of the negatively-charged-excitons [39]. Figure 2 (d) shows the power-dependent PL spectra of crystal-phase QDs at 4.2 K. The emission peak appears first at the high-energy side with narrow linewidth originating from the negatively-charged-exciton X$^-$. No fine structure splitting is observed indicating the occurrence of singly-charged states similar to those reported for heterojunction QDs [40]. The corresponding g-factor tensor and diamagnetic coefficient will be discussed in the following. A further increase of the excitation power leads to state-filling related emission that appears at low energy side, inducing the emission of negatively-charged-biexcitons XX⁻ with slightly larger linewidth with respect to X⁻. The emission PL intensity for X⁻ (XX⁻) as a function of excitation power is reported in right panel in Fig. 2 (e). It shows a nearly linear (quadratic) dependence that confirms the biexciton-exciton type transition [41]. The X⁻ emission shows a narrow linewidth of around 68 μeV (see left panel in Fig. 2 (e)), that is a clear indication of the good quality of crystal-phase QDs in our single InP NWs.

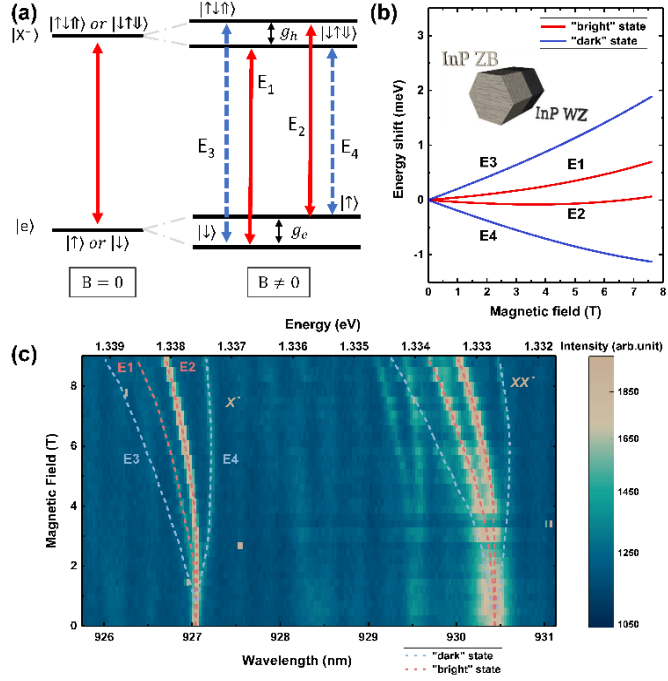

**Figure 3** (a) The energy diagram of the negatively-charged-exciton with the magnetic field perpendicular to the growth axis. Here $|\uparrow\rangle$ ($|\downarrow\rangle$) and $|\Uparrow\rangle$ ($|\Downarrow\rangle$) represent spin states of electron and hole, respectively. $g_e$ and $g_h$ are the electron and hole g-factors. (b) Illustration of the energy shift of the four possible transitions of charged exciton, as a function of the magnetic field. Red (blue) curves indicate bright (dark) transitions. The inset shows the simulated structures. (c) The magneto-PL spectra of X⁻ and XX⁻ with a magnetic field B$_{out-of-plane}$ from 0 to 9 T. The red and blue dashed curves represent the "bright" and "dark" states, respectively.

## 2.3 Out-of-plane magneto-PL spectroscopy

In the following, we discuss the changes in the charged-exciton energy levels of the present crystal-phase QDs under magnetic field. Figure 3 (a) shows the schematic diagram of negatively-charged-exciton energy levels in presence of an externally-applied magnetic field [42], where $|\uparrow\rangle$ ($|\downarrow\rangle$) and $|\Uparrow\rangle$ ($|\Downarrow\rangle$) represent the spin states of electron and hole, respectively. At zero magnetic field, the ground states $|e\rangle$ and charged exciton states $|X^-\rangle$ are degenerate [40, 43]: the single transition energy leads to a single PL peak. By switching ON the magnetic field, spin degeneracies are removed [40, 44] yielding the Zeeman splitting of $|e\rangle$ and $|X^-\rangle$ states governed by $g_e$ and $g_h$ [45,46]. The transitions $E_1(|\uparrow\downarrow\Downarrow\rangle \leftrightarrow |\downarrow\rangle)$ and $E_2(|\uparrow\downarrow\Uparrow\rangle \leftrightarrow |\uparrow\rangle)$ are shown by red

arrows in Fig. 3 (a) and satisfy the optical selection rule: they will be referred to as "bright" states. On the contrary, the transitions $E_3(|\uparrow\downarrow\Uparrow\rangle \leftrightarrow |\downarrow\rangle)$ and $E_4(|\uparrow\downarrow\Downarrow\rangle \leftrightarrow |\uparrow\rangle)$ (blue arrows) do not satisfy the selection rules and are thus referred to as "dark" exciton states. However, the presence of a magnetic field perpendicular to the NW growth axis makes the "dark" exciton states optically accessible, as a result of spin-eigenstate mixture [44]. The four transition energies ($E_1$, $E_2$, $E_3$, $E_4$) are determined to be [47]:

$$E_{i=1,2,3,4} = E_0 + \tfrac{1}{2}\mu_B\{\xi_e g_e + \xi_h g_h\}B + \gamma B^2$$

Here $E_0$ is the transition energy at zero-magnetic field, $\mu_B$ is the Bohr magneton, $\xi_{e,h} = (+,-)$ depends on the electron/hole spin orientation (different sign combination gives $E_i$), $g_e$ and $g_h$ are the electron and hole g-factors, while $\gamma$ represents the diamagnetic coefficient. According to Eq (1), the Zeeman splitting between two "bright" ("dark") states $E_1$ and $E_2$ ($E_3$ and $E_4$) depend on the summation (difference) of the $g_e$ and $g_h$. Therefore, by fitting the emission of the charged excitons as a function of the magnetic field B, the electron and hole g factor and diamagnetic coefficients can be extracted.

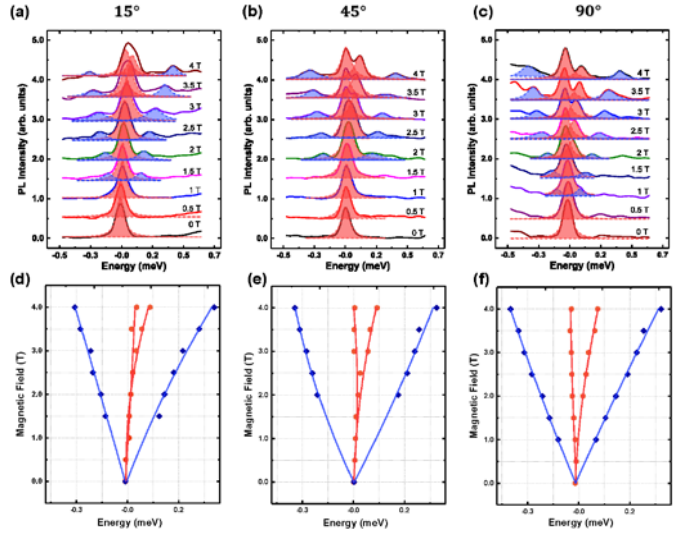

**Figure 4** (a)-(c) Magneto-PL mapping of the negatively-charged-exciton with in-plane magnetic field $B_{\text{in-plane}}$ from 0 to 4 T for three angles $\theta = 15°$, $45°$ and $90°$. The solid curves represent the measured magneto-PL spectra, and the blue (red) dashed lines with filled areas are the Lorentz fittings for the dark (bright) states. (d)-(f) Peak positions for different magnetic fields, extracted from (a)-(c), respectively. The blue dots and the red squares are the extracted data, and the blue and red solid curves are the fitted results through Eq. (1).

Figure 3 (b) illustrates the four possible energy transitions for a non-vanishing magnetic field. Specifically, we performed a 3D numerical simulation of the confined quantum states in the structure shown in the inset (i.e. a ZB/WZ QD), subject to a magnetic field orthogonal to the NW axis. An effective-mass two-band model was used, with a NW radius of 80 nm. Electron and hole levels were Zeeman split by using the experimental value of the g-factors. The thickness of the ZB and WZ regions for the spatially indirect transition, was taken equal to the exciton Bohr radius in the corresponding material.

A vertical magnetic field ($B_{\text{out-of-plane}}$, see Fig. 2 (a)) perpendicular to the NW axis was applied first and the magneto-PL measurements were performed. Figure 3 (c) shows the magneto-PL spectra and the evolution of X⁻ and XX⁻ emission energies ("bright" states correspond to the two inner peaks and are labeled by the red dashed lines, while the "dark" states are labeled by blue dashed lines for the outer transitions). Experimental data are nicely consistent with the theoretical calculation presented in Fig. 3 (b).

**2.4 In-plane angle-dependent magneto-PL spectroscopy**

Angle-dependent magneto-PL measurements under in-plane magnetic field ($B_{\text{in-plane}}$, see Fig. 2(a)) with different angle (θ) were performed to extract the g-factor tensor of WZ/ZB crystal-phase QDs in InP NWs. Solid curves in Figs 4 (a)-(c) report experimental magneto-PL spectra for three selected angles θ = 15°, 45°, and 90° (θ labels the angle between the in-plane magnetic field $B_{\text{in-plane}}$ and the NW axis). In the same figure, blue (red) dashed curves with filled area provide Lorentz fittings for the "dark" ("bright") exciton states. The fitted peak energies as a function of the in-plane magnetic field intensity are plotted in Figs 4(d)-(f). The splitting of both "bright" and the "dark" states increases significantly with θ changing from 15° to 90°. The different behavior of both these states with θ can be explained by taking into account the anisotropies of the g-factor tensor and diamagnetic coefficient in presence of an in-plane magnetic field. In fact, the variation of g-factor with θ can be described as [23, 47, 48]:

$$g_e(\theta) = \sqrt{\left(g_e^{[111]} \cos\theta\right)^2 + \left(g_e^{[11\bar{2}]} \sin\theta\right)^2} \quad (2),$$

$$g_h(\theta) = \sqrt{\left(g_h^{[111]} \cos\theta\right)^2 + \left(g_h^{[11\bar{2}]} \sin\theta\right)^2} \quad (3),$$

where [111] is the NW growth axis and [11$\bar{2}$] represents one of the axes of the NW facet. Figure 5 (a) reports the extracted $|g_e|$ and $|g_h|$ at different θ. The evolutions of electron and hole g-factors as function of θ were fitted through Eq. (2) and Eq. (3) (solid curves in Fig. 5 (a)) and from this we obtain $\left|g_e^{[11\bar{2}]}\right| = 2.07$ and $\left|g_e^{[111]}\right| = 1.84$, as well as the corresponding hole g-factor tensor $\left|g_h^{[11\bar{2}]}\right| = 1.35$ and $\left|g_h^{[111]}\right| = 0.44$. The hole g-factor shows a larger anisotropy with respect to the electron values, suggesting the hole g-factor sensitive to the geometry of the crystal–phase QDs in NWs.

The diamagnetic coefficient γ is largely determined by electron and hole wave function spatial extension according to $\gamma = \frac{e^2}{8m_\perp^*}\langle\rho^2\rangle$ [49, 50]. Here *e* is the electron charge, $m^*_\perp$ is the reduced mass of the exciton and can be derived from $\frac{1}{m^*_\perp} = \frac{1}{m^*_{e\perp}} + \frac{1}{m^*_{h\perp}}$, with $m^*_{e\perp}$ and $m^*_{h\perp}$ representing the electron and hole effective mass perpendicular to the applied magnetic field. The quantity $\langle \rho^2 \rangle$ represents the average extension of the wave function perpendicular to the applied magnetic field [49, 50]. The diamagnetic coefficient ranges from $\gamma \approx 4.5\ \mu eV\ T^{-2}$ at $\theta = 15°$ to $\gamma \approx 1.5\ \mu eV\ T^{-2}$ at $\theta = 90°$ (Fig. 5 (b)) confirming a pronounced system anisotropy.

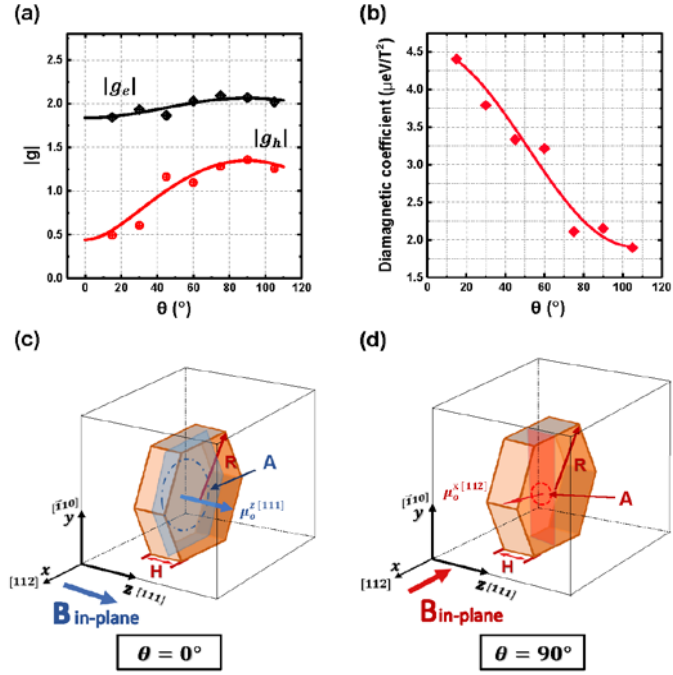

**Figure 5** (a) The absolute values of the electron and hole g-factor as a function of θ. The black squares and red circles are the measured data, and the black curve and red curve are the corresponding fitted results. (b) Diamagnetic coefficient as a function of θ with the red square and curve representing the measure data and the fitted results respectively. The schematic of the orbital moment with in-plane magnetic field with $\theta = 0°$ (c) and $\theta = 90°$ (d). Here x, y and z represent the [112], [$\bar{1}$10], and [111] direction, respectively. **A** is the area encircled by the orbital current $I_s$. **R** and **H** are are the radius and the height of the hexagonal prism. When the in-plane magnetic field (B$_{in\text{-}plane}$) is parallel to the NW axis, $\theta = 0°$, the electron and hole encircle in the plane (blue circle) of the NWs, introducing the z component (or [111]) of the orbital moment $\mu^z \propto R^2$. Instead, with the in-plane magnetic field (B$_{in\text{-}plane}$) perpendicular to the NW, $\theta = 90°$, the electron and hole confined in the narrow WZ/ZB segments, yielding a x component (or [112]) of orbital moment $\mu^x \propto H^2/4$.

## 2.5 Spin-correlated orbital current model

In order to qualitatively rationalize the g-factor and diamagnetic-coefficient anisotropies, the spin-correlated orbital-current model can be invoked [47]. The magnitudes of the spin and orbital moments $\mu_s$ and $\mu_o$ are proportional to the area **A** encircled by the spin and orbital currents $I_s$ and $I_o$. While the spin moment $\mu_s$ can be assumed isotropic and close to $\mu_B$ [47, 52], the orbital moment $\mu_o$ is instead dependent on the shape of the nanostructure.[51] For the WZ/ZB segments of height **H** and characteristics radial dimension **R** (see Fig. 5 (c) and (d)), the orbital moment involves different components. When θ = 0°, electron and hole encircle the blue area in Fig. 5 (c) thus yielding a z ([111]) component of the orbital moment $\mu_o^z \propto R^2$. At θ = 90°, the electron and hole encircle the red area in Fig. 5 (d): here the x ([112]) component $\mu_o^x$ dominates. Differently from $\mu_o^z$, $\mu_o^x \propto \left(H/2\right)^2$ results in smaller $\mu_o^x$ than $\mu_o^z$.

According to the Zeeman interaction and time-reversal symmetry, the g-factor tensor is determined by $g_{e,h}^{x,y,z} = 2\{\mu_B(\mu_s^{x,y,z} + \mu_o^{x,y,z})\}^{-1}$. Due to the spin moment $\mu_s^{x,y,z} \sim \mu_B$, the orbital moment in different directions likely reflects the g-factor. For WZ/ZB segments with large radius and small height, $\mu_o^x < \mu_o^z$ results in $g^x > g^z$, which is consistent with the experimental results of $\left|g_e^{x([112])}\right| = 2.07$ and $\left|g_e^{z([111])}\right| = 1.84$ ($\left|g_h^{x([112])}\right| = 1.35$ and $\left|g_h^{z([111])}\right| = 0.44$).

The diamagnetic coefficient γ, related to the average extension of the wave function $\langle \rho^2 \rangle$, is sensitive to the aspect ratio of the nanostructure. The wave function at θ = 0° is more delocalized than the wave function at θ = 90°, thus γ (θ = 0°) is larger than γ (θ = 90°) as observed in the experimental data in Fig. 5 (b). This dependence of electron/hole g-factor tensor and exciton diamagnetic coefficient in WZ/ZB QDs on the geometry of the nanostructure can be exploited to control spin properties by tuning the g-factor tensor via suitable choice of the WZ and ZB segment geometry.

## 3  Conclusion

In conclusion, the anisotropy of electron and hole g-factor tensor and of the exciton diamagnetic coefficient in WZ/ZB crystal-phase QDs realized in InP NWs have been investigated by angle-dependent magneto-PL measurements at 4.2 K. The electron and hole g-factor tensor and exciton diamagnetic coefficient were assessed experimentally and correlated to the structure aspect ratio. The spin-correlated orbital current model allowed us to rationalize the experimental results. While the quantum-confinement depending on the size of the crystal phase QDs allows us to observe PL emission peaks in the range from 1.27 to 1.35 eV, we investigated the g-factor and diamagnetic shift from each isolated peaks, and our results are not affected by the size distribution of the crystal phase QD that may occur in the investigated NW. Our findings provide useful information on crystal-phase structure in InP NWs, and indicate a strategy to control the electron spin in crystal-phase structure by tuning the g-factor tensor. Since the interface of the crystal phase QDs can go to atomic level, it provides a great potential to control the band structure precisely, thereby the electronic and optical properties. This makes the crystal phase QD system have applications in photosensors, solar cells and lasers. Meanwhile, the high controllability of crystal phase in NWs enables the manipulation of the quantum state coupling and entanglement, which broadens the selectivity for implementing the applications in quantum information processing and spintronics.

## 4  Experimental Section

InP NWs were grown by chemical beam epitaxy with an Au-assisted approach on a [111]-oriented InP substrate. Tapered nanowires with hexagonal cross-section and [112]-oriented side facets were obtained. The NW structures were characterized via the HAADF images using a STEM, in order to identify the ZB and WZ segment distribution. Then NWs were deposited on the Si/SiO$_2$ substrate patterned with metallic markers which allowed us to identify the position and the orientation of the individual NWs. The latter were characterized by micro-PL. The schematic diagram of the micro-PL setup is shown in Fig. 2(a). The samples were mounted on a xyz piezoelectric platform controllable with 50 nm accuracy in a low temperature (4.2 K) cryostat. Two Nb/Ti orthogonal superconducting split-coils were used to apply magnetic fields at different angle. The confocal PL microscope was furnished with an objective with a large numerical aperture of NA=0.82. Non-resonant excitation was obtained by a 650-nm semiconductor laser for micro-PL measurement. The PL signals were dispersed by a 0.55-m spectrometer and collected via a liquid nitrogen cooled charge coupled device (CCD) camera with a spectral resolution of about 60 μeV. A LED illumination and a complementary metal oxide semiconductor (CMOS) camera were used to monitor position and orientation of the NWs. Two different magnetic field configurations were employed, as sketched in Fig. 2 (a). Here we define κ as the excitation (collection) direction and a as the NW axis. The orange arrow label with Bout-of-plane represents the out-of-plane magnetic field configuration with the magnetic field B parallel to the pumping direction κ. The blue arrow label with Bin-plane represents the in-plane magnetic field configuration with magnetic field B perpendicular to κ. In both configurations, the direction of the excitation laser κ is perpendicular to the NW axis a. In the in-plane configuration, the angle between the magnetic field and the NW growth axis a is denoted with θ.

The numerical simulations of the PL peaks diamagnetic shift were performed by computing the single-particle electron and hole confined energies by means of a box integration method in the 3D domains shown in the insets of Fig. 3 (b) [53], including an orthogonal magnetic field. An effective-mass two-band model was used, with a NW radius of 80 nm and the following InP effective masses: $m_e^* = 0.064$, $m_{h[100]}^* = 0.52$, $m_{h[111]}^* = 0.95$ for the ZB region, and $m_{e[1000]}^* = 0.073$, $m_{e[0001]}^* = 0.064$, $m_{h[1000]}^* = 0.14$, $m_{h[0001]}^* = 1.25$ for the WZ region [37]. The conduction and valence band energy offsets were taken as 129 meV and 45 meV, respectively. The thickness of the ZB and WZ regions for the spatially indirect recombination were taken equal to the exciton Bohr radius in the corresponding material [54], namely 11 nm and

9 nm.

**Acknowledgements**

This work was supported by the National Natural Science Foundation of China under Grants No. 11934019, No. 61675228, No. 11721404, No. 51761145104 and No. 11874419; the Strategic Priority Research Program, the Instrument Developing Project and the Interdisciplinary Innovation Team of the Chinese Academy of Sciences under Grants No. XDB28000000 and No.YJKYYQ20180036, the Key RD Program of Guangdong Province (Grant No.2018B030329001), and the Key Laboratory Fund under (No. 614280303051701). We acknowledge financial support from the SUPERTOP project, QUANTERA ERA-NET Cofund in Quantum Technologies.